# Intermittent Signals and Planetary Days in SETI

**Robert H. Gray**
Gray Consulting, 3071 Palmer Square, Chicago, IL 60647, USA
email: RobertHansenGray@gmail.com  orcid 0000-0002-4911-5832

**Abstract:** Interstellar signals might be intermittent for many reasons, such as targeted sequential transmissions, or isotropic broadcasts that are not on continuously, or many other reasons. The time interval between such signals would be important, because searchers would need to observe for long enough to achieve an initial detection and possibly determine a period. This article suggests that: (1) the power requirements of interstellar transmissions could be reduced by orders of magnitude by strategies that would result in intermittent signals, and (2) planetary rotation might constrain some transmissions to be intermittent and in some cases to have the period of the source planet, and (3) signals constrained by planetary rotation might often have a cadence in the range of 10-25 hours, if the majority of planets in our solar system are taken as a guide. Extended observations might be needed to detect intermittent signals and are rarely used in SETI but are feasible, and seem appropriate when observing large concentrations of stars or following up on good candidate signals.

## 1. Introduction

Searches for technosignatures of extraterrestrial intelligence (SETI) usually require that any signal be present all or much of the time in order to be detected, because observations are usually brief—a few minutes in the case of major transit surveys such as Ohio State (Dixon 1985), META (Horowitz & Sagan 1993), META II (Colomb et al. 1995) and BETA (Leigh & Horowitz 2000), or a fraction of an hour in the case of targeted searches such as Phoenix (Backus et al. 2004), ATA (Harp et al. 2016) and *Breakthrough Listen* (Enriquez et al. 2018). But signals might be intermittent for many reasons (Shostak 2011a, Shostak 2009, Benfords 2010, Sullivan 1991) such as isotropic transmissions with duty cycle <1 or targeted sequential transmissions. Most searches to date (Tarter 1995, Tarter 2001, and updates) would be unlikely to detect signals with repetition rates of many hours because most have not dwelled very long on targets and the population of signals is presumably small.

Follow-up searches for candidate signals are also usually brief, and consequently might miss intermittent signals that happen to have been detected one time. For example, nine META candidate positions were re-observed for 5-10 minutes each (Lazio, Tarter & Backus 2002), and 226 candidate positions from SETI@home and the SERENDIP project at Arecibo were re-observed (Korpela, Cobb, Werthimer & Lebofsky 2004) for approximately 13 s each (Korpela, E., private communication with Gray). The Ohio State 'Wow' candidate signal (Kraus 1994) is a rare exception, with over 100 hours of follow-up observing time (Harp et al. 2020, Gray & Ellingsen 2002, Gray & Marvel 2001, Gray 1994).

This article suggests that:

- Some signals might be intermittent as a consequence of the large distances involved in interstellar signaling. Continuous isotropic radio transmissions would require an enormous amount of power, and the average power could be reduced by orders of magnitude by the simple expedient of a lower duty cycle resulting in an intermittent signal; highly directive radio or optical transmissions could also greatly reduce power requirements but would appear intermittent if targets are illuminated sequentially.
- Some signals might be intermittent as a consequence of planetary rotation, a ubiquitous mechanism that would affect both transmissions and observations from the surface of planets. Transmissions from a single site could be interrupted for some part of each day for targets that are not always above the horizon, and in the special case of a fixed directional antenna system, the transmission might sweep across distant observers once each 'day' periodically like a lighthouse.
- Days might often be in the range of 10-25 hours, as are the majority of our planets, if the distribution of planetary days elsewhere is comparable to our solar system.
- Searches might require extended observations to detect intermittent signals, potentially for periods of time comparable to days, or observe a sufficiently large number of targets briefly to get a detection by chance.

Reports of one-time detections of candidate signals (for example, Horowitz & Sagan 1993, Kraus 1994) also suggest that follow-up observations should be extended in time to accommodate the possibility that the events might repeat as was the case of FRB 121102 (Spitler et al. 2016).

**2. Intermittent Signals**

*2.1 Reasons to anticipate intermittent signals*

Isotropic radio broadcasts are often assumed or used as examples in discussions of interstellar signaling, for example "... we feel the practicality of beamed transmission begins to disappear at about 50 light years and vanishes for $r >$ 100 light years. There is then no alternative to broadcast transmission." (Oliver 1993). Many other examples exist, such as Tingay et al. 2016 and Oliver & Billingham 1971. But, isotropic broadcasts require enormous amounts of power if radiating continuously (Shostak 2011b, Shostak 2009, Pfleiderer 1988).

One reason to consider the possibility of intermittent signals is that the average power required for isotropic broadcasts could be reduced by many orders of magnitude by the simple expedient of reducing the duty cycle by comparable factors. Another reason is that high-gain transmissions provide an alternative requiring much less power than continuous isotropic (Shostak 2007) but would illuminate many fewer potential observers in the case of one or a small number of beams directed at targets sequentially; optical transmissions directed by large telescopes might similarly be intermittent due to sequential pointing. These two intrinsically intermittent cases are discussed in the next two sections primarily in the context of radio, followed by discussion of several other cases where signals might merely seem intermittent but could be detected by more sensitive or more sophisticated searches.

*2.2 Isotropic broadcasts*

Speculation about the distance to communicative civilizations often fall in the range of $\sim 10^2$-$10^3$ ly (for example, Ekers et al. 2002 p. 115) so those ranges are used in the following examples.

Table 1 shows that an isotropic transmission with a $10^3$ ly range would require $\sim 10^{15}$ W to produce a signal-to-noise ratio SNR = 10 in 1 s, assuming a receiver system with a 100-m antenna comparable to some current searches (e.g. Enriquez et al. 2018) and using formulae from Gray & Mooley 2017. That exceeds total current terrestrial power consumption of $\sim 10^{13}$ W (BP 2019) by orders of magnitude and it is $\sim 1\%$ of total terrestrial $\sim 10^{17}$ W insolation (Coddington et al. 2016) which might raise environmental impact issues if conducted on the surface of a planet (Rebane 1993). Shorter ranges still require a great deal of power with isotropic broadcasting, even though power required decreases with $1/r^2$; reducing the range to 100 ly encompassing $\sim 10^3$ stars requires $\sim 10^{13}$ W or $\sim 10^4$ 1,000 MW power plants.

These examples far exceed the capability of civilizations comparable to ours. The large power required for continuous isotropic broadcasts could conceivably be available to some very technologically advanced civilizations (Kardashev 1964, 1967), but assuming very advanced civilizations seems very optimistic. The 1 Hz channel width in this example was selected to minimize power requirements, but implies a data rate of ~1 bit per second which is very slow compared with common data rates such as $\sim 10^4$ bps (audio) to $\sim 10^6$ bps (video) which would require orders of magnitude more power.

| | | | | | | |
|---|---|---|---|---|---|---|
| Table 1 ||||||||
| Power Required for 1,000 Light Year Range at 1.42 GHz |||||||
| Transmitting Antenna |||||||
| Example antenna: | Isotropic | META | GBT | Arecibo | None available | None available |
| Diameter (m) | n/a | 30 | 100 | 300 | 1,000 | 3,000 |
| Gain (approx.) | 1 | $10^5$ | $10^6$ | $10^7$ | $10^8$ | $10^9$ |
| Power, W (approx.) | $10^{15}$ | $10^{10}$ | $10^9$ | $10^8$ | $10^7$ | $10^6$ |
| Example power source | >terrestrial power generation | biggest power plants | big power plant | small power plant | locomotive | wind turbine |
| Notes. Receiver assumptions: antenna diameter=100 m, efficiency=0.7, frequency=1.42 GHz, $T_{sys}$=20 K, channel bandwidth=1 Hz, integration time=1 s, SNR=10. Transmitter assumptions include antenna efficiency=0.7 except for isotropic where gain=1.0, and single-aperture primary beamwidth. ||||||||

Broadcasters could reduce the average power requirements of isotropic transmissions by many orders of magnitude with simple strategies such as reducing duty cycle by comparable factors. For example, reducing the duty cycle to 1% could provide a 100-fold reduction in average power required, perhaps radiating for 1 s out of every 100 s. Searches observing targets for a matter of minutes might detect such signals, such as the Ohio State and META transit surveys which observed objects for 72 s and 120 s respectively, or *Breakthrough Listen* observing targets for three five minute periods (Enriquez et. al. 2018), or a targeted search such as Phoenix observing objects for 1,000 s in each of several spectral windows (Cullers 2000), or the ATA observing for 30 minutes (Harp et al. 2016). Reducing duty cycle further yields further savings—for example a $10^{-4}$ duty cycle with a $10^4$ reduction in average power might result in a 1 s signal every three hours, but most searches to date would be likely to miss such signals. Assuming longer signal duration does not help much; a 1-hour signal present every 100 or 10,000 hours would be very unlikely to be found by most current search strategies unless the population of such signals is large.

There are no obvious and unique time intervals that strongly suggest some specific duty cycle or signal duration, although some possibilities are noted later.

*2.3 Targeted transmissions*

Another way to reduce transmission power requirements is to use high-gain antenna systems to direct power at targets such as single stars (Shostak 2007). For example, Table 1 shows that with an Arecibo-scale transmitting antenna the power requirement falls to familiar levels: $10^8$ W or one-tenth of a big power plant for a 1,000 ly range which is a $10^7$ reduction compared with isotropic due to the $10^7$ gain. For a 100 ly range, the power falls to $10^6$ W which is the rating of a planetary radar at the Arecibo observatory (Campbell, Hudson, & Margot 2002).

If a single beam is directed at targets in succession, the signal would appear intermittent to observers. Many beams might illuminate many targets simultaneously, but the total power required increases with the number of targets. The number of stars within various ranges in the Galactic plane has been estimated as approximately $n = (r/8)^3$ where $r$ is range in light years (Drake, Wolfe & Seeger 1984), which results in $\sim 10^6$ stars within 1,000 ly, $\sim 10^3$ within 100 ly, and $\sim 1$ within 10 ly ($\sim 10$ is more nearly consistent with observations).

In one example assuming a 64 m antenna on each side (Drake, Wolfe & Seeger 1984), 2,000 targets out to 100 ly could be constantly illuminated by 2,000 high-gain beams using a total of 900 MW, which is comparable to one big power plant. Increasing the number of targets and beams increases the power required until an isotropic broadcast would be more efficient at 950 ly; illuminating 2 million targets would require a total of $6.8 \times 10^{13}$ W. The power requirements differ from Table 1 because the ranges and system assumptions in this example are somewhat different.

Antenna systems are conceivable that would form extremely narrow beams to illuminate distant planetary systems more efficiently than typical radio telescopes (Shostak 2011b) greatly reducing power requirements. Large optical telescopes have been proposed to similarly illuminate planetary systems efficiently with lasers (Howard et al. 2004).

A targeted transmission with one beam illuminating targets in succession and repeatedly would appear intermittent to observers, with an effective duty cycle depending largely on the number of targets and dwell time on each one. There is no obvious way to estimate how often such signals might repeat or their duration, but if the number of targets is in the range of $10^3$-$10^6$ stars within 100-1,000 ly, the effective duty cycle could be $10^{-3}$-$10^{-6}$ for a single beam. It seems very optimistic to assume that we are a target all or most of the time, which would be required for our typically brief observations to detect a targeted sequential transmission.

*2.4 Variable Detectability*

The two possible sources of intermittency considered so far—isotropic transmissions with duty cycle <1 and targeted sequential transmissions—would be intrinsically undetectable at some times; increasing search system sensitivity would not increase the chances of detection when we are not illuminated. Several other types of signal might merely seem intermittent but could be detected by more sensitive or sophisticated searches, as noted in following sections.

*2.4.1 Variable Flux*

Some signals might appear to be intermittent due to variations in power at the source or propagation effects.

Radiated power might be intentionally varied for many reasons. For example, occasionally increasing the power by a factor of four would increase range by a factor of $4^{1/2} = 2$ and the total number of uniformly distributed potential recipients would increase by a factor of $2^3 = 8$. Such signals might appear to be intermittent if the increased flux exceeds the receiving system detection threshold, but at other times falls below the threshold.

Another possible cause of variable flux might be variable power cost or availability, which could result in a diurnal variation in flux for transmissions from planets if power is cheaper at night (often the case with terrestrial generation), or cheaper during daylight (due to photovoltaic production). If photovoltaic is the only source of power (and neither stored nor distributed over long distance), flux might be absent during night time.

Interstellar scintillation could sometimes boost a usually undetectable signal above a detection threshold, or attenuate an otherwise detectable signal below the threshold (Cordes & Lazio 1991). Scintillation gain depends on mainly on frequency, direction, and distance (Cordes, Lazio, & Sagan 1997) and the effect can occasionally be large (>10) for monochromatic signals over long distances (~100 pc) at centimeter wavelengths (Cordes, Lazio, & Sagan 1997). The scintillation timescale is on the order of seconds to hours over distances of ~10 kpc, and one strategy for accommodating varying scintillation gain is multiple observations at separate times rather than a single observation (Cordes & Lazio 1991).

*2.4.2 Variable Frequency*

Frequency is often presumed constant, or slowly changing due to orbital and rotational Doppler effects that require de-drifting in narrowband searches of rates that could be as high as 200 Hz s$^{-1}$ at 1 GHz (Sheikh et al. 2019), but frequency might vary much more for other reasons. For example, transmission frequency might occasionally be changed to accommodate observers using different spectral windows. Many SETI surveys have been near 1.42 GHz using relatively narrow ~1 MHz spectral windows (Tarter 1995, Horowitz & Sagan 1993, Dixon 1985) which covers only

~$10^{-4}$ of the terrestrial microwave window 1-10 GHz (Oliver & Billingham 1971 p. 41), yet many other 'magic' frequencies have been proposed, and senders might use different frequencies at different times. A few searches have covered wider spectral windows such as BETA with 1.4-1.7 GHz (Leigh & Horowitz 2000) and some recent searches such as SERENDIP (Chennamangalam et al. 2017), *Breakthrough Listen,* and the ATA are now covering multiple GHz typically in steps rather than simultaneously.

*2.4.3 Variable Polarization*

Signal polarization might vary, for example between left and right circular as a form of modulation (Dixon 1973), which could make a signal appear intermittent to a receiver sensitive to only one polarization. Some SETI experiments process and analyze two orthogonal polarizations separately to accommodate varying or intrinsically polarized signals (e.g. Horowitz & Sagan 1993; Gray & Ellingsen 2002), while others analyze total power (Stokes I) reducing sensitivity to an intrinsically polarized signal (e.g. Siemion et al. 2013; Enriquez et al. 2018).

*2.4.4 Variable Bandwidth*

Signal bandwidth might vary over time for several reasons, such as transmissions sometimes using a narrowband signal as a 'beacon' to maximize the range of detectability and at other times using the same power to encode information using a wider bandwidth, and for search systems with fixed channel widths, such signals might appear to be intermittent—detectable when signal and receiver channels approximately match, but not at other times. SETI@home (Korpela et al. 2011) is an example of analyzing many different channel widths and drift rates attempting to match unknown signal characteristics.

*2.5 Time Intervals in SETI*

An estimate of how often intermittent signals might appear and their duration would be useful in designing search strategies and follow-up searches. Repetition rates of seconds or minutes could be detected by current search strategies which observe that long, but longer intervals which might result from low-duty-cycle isotropic broadcasts or targeted transmissions or other sources of variability would likely be missed. There are no obvious unique time intervals that signalers might be constrained to use or choose to use, but planetary rotation might affect transmissions from the surface of planets, which might make the length of planetary days a factor in SETI—making some signals intermittent and possibly making some periodic.

# 3. Planetary Days

*3.1 Planetary Days in the Solar System*

The planets in our planetary system provide a sample for estimating rotation periods in other planetary systems. Table 2 shows the current day and selected statistics for nine planets (Pluto is included although it was reclassified as a dwarf planet by the IAU in 2006; results are presented with and without Pluto). The median day is 23.9 hours; if Pluto is excluded, the median day is 21.1 hours (a mean of Neptune and Earth). The table also presents mean days, some of which are much longer, but medians seem more appropriate because means are sensitive to extreme values such as Mercury and Venus at >1,000 hours, and the number of cases is small.

| Table 2 | | | | |
|---|---|---|---|---|
| Planetary Days in the Solar System | | | | |
| Planet | Sidereal period (hours) | Extremes excluded | Rocky | In solar HZ | Life known |
| Jupiter | 9.9 | | | | |
| Saturn | 10.5 | 10.5 | | | |
| Uranus | 15.6 | 15.6 | | | |
| Neptune | 18.4 | 18.4 | | | |
| Earth | 23.9 | 23.9 | 23.9 | 23.9 | 23.9 |
| Mars | 24.6 | 24.6 | 24.6 | 24.6 | |
| Pluto | 152.9 | 152.9 | 152.9 | | |
| Mercury | 1403.7 | 1403.7 | 1403.7 | | |
| Venus | 5816.3 | | 5816.3 | | |
| N | 9 | 7 | 5 | 2 | 1 |
| Mean | 830.6 | 235.6 | 1484.3 | 24.2 | 23.9 |
| Median | 23.9 | 23.9 | 152.9 | 24.2 | 23.9 |
| Reference: Zombeck 2007 p. 51. | | | | | |

Two-thirds of nine planets have days in the range of approximately 10-25 hours, and the middle third are in the range of 18-25 hours. If Pluto is excluded, three-quarters of the eight planets have days in the range of approximately 10-25 hours, and the middle half are in the range of 15-25 hours. Median days for subsets that might be most relevant for planets with life are: 1484 hours for rocky planets (five cases), 24.2 hours for planets in or near the current solar habitable zone 0.95-1.67 AU (Kopparapu et al. 2013, Kasting et al. 1993) (two cases), and 23.9 hours for the one planet known to have life. The days of rocky planets are clearly more relevant than those of gas giants, where neither transmitters nor receivers would be expected.

Some days have varied considerably over time. Mercury is thought to have been trapped in a spin–orbit 3:2 resonance for most of its history (Noyelles et al. 2014), and Venus is thought to have 'spun down' from a period of possibly several days due in part to its massive atmosphere (Correia & Laskar 2003). The Earth is known to have spun down due lunar-solar tidal friction, with the terrestrial day estimated as 21.9±0.4 hours 620 million years ago based on analysis of sedimentary tidal rythmites (Williams 2000), and ~18 hours 900 million years ago (Sonett et al. 1996).

Large samples of exoplanet days may become available in the future, but few are currently available, and selection effects in detection may make unbiased samples difficult to get. The first exoplanet day reported was the young planet β Pictoris b with a day estimated as 8.1±1.0 hours (Snellen et al. 2014). One estimate of the largest rotation rate that a planet like Earth can have with out breaking up is about 84.5 minutes (Sheikh et al. 2019). One simple model of formation focused on planets with masses less than 10 times the Earth run over $2 \times 10^7$ yr (Miguel & Brunini 2010) found that for simulations starting from 1,000 different discs, rotation periods for primordial planets ranged from ~10 h up to nearly 10,000 h with an approximately flat distribution, with some between 0.1 and 10 h.

The statistics for our planetary system suggest that many planets in other systems might have periods in the range of 10-25 hours, and that might be useful information in searching for interstellar signals, because the length of the day might affect the repetition rate of some signals.

*3.2 Effects of Planetary Days on Interstellar Signaling*

Planetary rotation might affect both transmissions and searches in several ways discussed in the following sections.

*3.2.1 Interstellar Lighthouse*

In one special case, a directional transmission from a fixed antenna system on a rotating planet could result in an intermittent signal having the period of the planet's sidereal day. Such a transmission might appear like a lighthouse—bright for a short time as it swept across an observer, absent for the rest of the source planet day, and repeating with the planet's rotation period.

An interstellar lighthouse could result from a fixed antenna system illuminating a spherical lune from north to south poles, or with phased array beams scanning from north to south poles, or with other approaches. Illuminating a $1^o$ lune, for example, would require 1/360 of the power of an isotropic broadcast which is a substantial savings. In the case of a source planet with the median day in our planetary system and a rotating $1^o$ lune, distant observers would be illuminated for 23.9/360 = 0.0664 hours or 3.9 minutes every 23.9 hours. Such a signaling strategy would have the isotropic broadcasting advantage of illuminating the entire sky although not constantly, and the directional transmission advantage of much lower power requirements than isotropic, and with no need for tracking. A transmission from a rotating antenna system might display a signature Gaussian rise and fall as it swept across a detector, and that might suggest re-observation efforts scaled to a planetary day.

*3.2.2 Shadowing*

A more general effect of planetary rotation would be periodic shadowing of some transmissions and observations in operations at single sites on the surface of planets. Targeted transmissions would not be possible when potential targets are below the horizon, so observers could see no signal for some part of each sender's day.

A single site can limit operations to much less than one-half day. For example, a transmission from a terrestrial site at the latitude of the Very Large Array ($34^o$ north) toward a target at declination of $-27^o$ would be possible during only the four hours each day during which targets are above the horizon, and impossible for the other 20 hours when targets would be shadowed by the Earth.

Observers from a single site on the surface of a planet have a similar problem with some targets sometimes being shadowed by their planet. Optical observations have the additional constraint of usually being conducted at night when shadowed from sunlight. This shadowing effect of planetary rotation is a reason to consider the possibility of a daily cadence in both radio and optical SETI.

*3.3 Effects of Planetary Days Can Be Avoided*

Planetary rotation might affect many transmitters and observers, but not necessarily all. Not all stars rise and set from the perspective of a site located away from an equator, and half of the stars never set for a site at a pole; in the case of tidally locked planets, rotation would be very slow. Multiple scattered sites such as the Deep Space Network (JPL 2000) can be used to illuminate or observe any celestial position at any time, although increasing the cost of operations. Operations from satellites or other spacecraft could reduce or eliminate the effects of planetary rotation, although further increasing cost and complexity. Most of our searches and transmissions have been conducted from single sites on the surface of a planet, so the single-planetary-site scenario and its consequences seem worth considering.

**4. Other Time Intervals**

Planetary days seem especially relevant in SETI, because both transmitters and observers located on planets would be affected by diurnal rotation, and because planetary days would be a widely known time scale. But, other time intervals might be relevant, and some are reviewed in the following sections.

*Pulse periodicities*. Sullivan suggested that "more attention should be paid in SETI programs to the possibility of finding rationalized, preferred pulse periodicities, in the same sense that many have argued for preferred frequencies" (Sullivan 1991) and listed 21 possible time intervals—many in a 'pulse window' between 0.1 s and 3.0 s defined in part by the distribution of pulsar periods which presumably would be widely known—and he noted that range of time intervals also includes many terrestrial heart beat rates. The longest pulse window presented was the terrestrial day.

*Terrestrial day*. The terrestrial sidereal day has been suggested as a period that some observers might detect due to our transmissions from fixed antennas with radiation patterns that sweep across the sky (Sullivan et al. 1978), and it has been suggested that the same time interval might be used in a signal sent back to us (Sullivan 1991). In the case of optical signaling, it has been suggested that targets might be illuminated for about twice the terrestrial day or 50 hours in order to have a good chance of arriving at night—taking that as the "longest likely rotation time of livable planets" (Ross 2000).

   *Years*. Orbital periods are known for many exoplanets (Han et al. 2014), so the year seems worth considering as a time interval in interstellar signaling. It has been suggested that signalers on exoplanets could use laser emissions to modify their transit profile in a clearly artificial way or transmit information during their transit to attract the attention of observers (Kipping & Teachey 2016), which would make each exoplanet year a special time interval for distant observers. Years do not, however, offer a general time interval that would be useful in SETI, because they vary so much. In our solar system, years range from 0.24 Julian years for Mercury to 163.73 for Neptune (Zombeck 2007 p. 52), and in a sample of 2,950 confirmed exoplanets the orbital periods vary from hours to centuries (Han et al. 2014). The year for planets with life might be less variable, but only one case is currently known. One estimate from fundamental physical constants for maintaining a life-supporting environment on a planet orbiting a star is ~2 years (Lightman 1984).

   *Terrestrial year.* For the special case of distant observers detecting the Earth transiting the Sun, it has been suggested that signals might be timed to arrive during our transit (Heller & Pudritz 2016; Castellano et al. 2004; Shostak & Villard 2002), which would make the terrestrial year a special time interval for observing in some directions. If signals are synchronized to arrive at Earth during our transit, we might detect them if we observe in the anti-Sun direction in the Earth's transit zone. An approximately 0.5 day transit once per year would require signaling in our direction with a duty cycle of 1/(365/0.5) or approximately 0.14% of the time, which would be a substantial savings for the transmitting side and for our search activities. The Earth's transit zone is only approximately 0.5° wide (Heller & Pudritz 2016) making this an inherently restricted case, and requires signalers to know the range with extreme accuracy.

   *Moons*. Transmitting from or searching from natural satellites has been suggested, with advantages including reduced radio frequency interference on a far side (Maccone 2019). Disadvantages include cost of transportation and operating in a likely hostile environment. Operations from moons would also be subject to rotation effects, likely often tidally locked with their planets. The seven largest moons in the solar system are all tidally locked, with rotation periods from 42 hours (Io) to 654 hours (Moon) and a median of 177 hours (Zombeck 2007 p. 33).

   *Synchronization*. Astronomical events have been suggested for coordinating the timing of transmissions and observations. A number of schemes involve exoplanets, ephemeris, and transits (Wright 2017). One scheme uses exoplanet ephemeris to forecast events such as the conjunction of two exoplanets along a line of sight from the Earth, aiming to intercept a transmission from the far planet toward the planet in between (Siemion 2014). Other schemes use synchronizing events such as nova (Makovetskii 1980), binary star mergers in other galaxies (Nishino & Seto 2018), or opposition of planets in different planetary systems (Corbet 2003).

   Compared with the various time intervals and synchronization schemes noted above, planetary days are an approximate time interval which might be useful in searching for interstellar signals because days would have a direct physical effect on many operations and might be widely known as often being in the range of several tens of hours.

## 5. Searching for Intermittent Signals

   If a large population of intermittent signals exists, typically brief SETI observations might achieve a detection by chance. An example from astronomy is the detection of four FRBs by the High Time Resolution Universe survey observing 4500 square degrees with 270 s pointings, resulting in an event rate estimate of approximately $10^4$ sky$^{-1}$ day$^{-1}$ (Thornton et al. 2013) even though most bursts appear to be one-time events. But, the population of detectable interstellar signals is presumably small and the energy and range much smaller than astrophysical events. One estimate of the population of interstellar signals is $10^5$ cases in the galaxy (Drake 1980), and the range of our searches is often cited as $10^3$ ly or $10^{-3}$ of the volume of the galaxy which might then contain $10^5/10^3 = 10^2$ cases among $10^6$ stars or one in $10^4$. If approximately 1,000 stars are observed for 15 minutes each per year as was the case with recent searches (Enriquez et al. 2018; Price et al. 2020), it might take ten years to eventually observe the one star in 10,000 where a transmission is always present—but if it is present for only 15 minutes during a 24 hour day, the chances of detection on that occasion are only about 0.01 (15/1440). Observing targets for say 24 hours rather than 15 minutes would reduce search speed by a factor of roughly 100 in the case of single-target observations (ignoring overhead), a disadvantage that could be eliminated by wide-field observations encompassing 100 targets simultaneously and with the potential advantage of greater sensitivity due to longer integration time—up to $100^{1/2} = 10$ greater assuming compensation for unknown Doppler drift rates.

   Observing many single stars or small fields for many hours or days searching for intermittent signals would be inefficient, but extended observations are feasible for selected targets and good candidates, and when many stars are observed simultaneously with wide-field techniques such as synthesis imaging. Repeating fast radio bursts are a recent example of discovery resulting from extended observations, where what had appeared to be one-time events scattered across the sky were found to repeat in the case of FRB 121102 by using three hours of observations (Spitler et al. 2016) and localized using 83 hours of observations (Chatterjee et al. 2017).

   Observing a single target for many hours is very rare in searches for technosignatures, but is common in radio astronomy in order to increase sensitivity, or to improve sampling of the UV plane, or for other reasons. In one extreme case, a 14-meter radio telescope at Mt. Pleasant has tracked the Vela pulsar for 18 hours per day for over 24 years totaling ~$10^4$ hours (Dodson et al. 2007). The most extended modern SETI observations of a single field to date is 100 hours, imaging a 2.5° field with the Allen Telescope Array (Harp 2019 submitted). The next most extended appears to be the first SETI project, in 1960 which observed two stars for approximately 100 hours each while stepping a single 100 Hz channel across 260 kHz (Drake 1985).

Future searches for interstellar signals and astrophysical transients may eventually monitor very wide fields or the entire sky. Some relatively wide-field aperture array radio telescopes are now operating at <300 MHz (Garrett et al. 2017) such as LOFAR (Low-Frequency Array), MWA (Murchison Widefield Array), and LWA (Long Wavelength Array), and the Ohio Argus prototype system which monitors a near hemisphere continuously (Ellingson et al. 2008) with a sensitivity of 66 kJy detecting the radio Sun and satellites. A 'fly's eye' technique has been demonstrated at L-band using 30 6-m ATA antennas pointing in different directions to monitor a 147 deg$^2$ field with 128 channels and 209 MHz spectral window (Siemion et al. 2012). One assessment of future prospects is that "An array of 1024x1024 half-wavelength dipoles would have a collecting area equivalent to an 85 m dish at 1.4 GHz and would require ~$10^{21}$ ops of computer power to tessellate all the sky above the horizon." (Tarter 2001 p. 542).

## 6. Conclusions

Some interstellar signals might be intermittent, for many possible reasons. One reason is that the enormous power required for continuous isotropic transmission could be reduced by orders of magnitude by the simple expedient of reducing the signal duty cycle by similar factors, or by using high-gain antenna systems or optical telescopes to illuminate targets sequentially. Another reason for intermittency could be targets of transmissions sometimes being below the local horizon, and another could be lighthouse-like transmissions from fixed antenna systems on rotating planets or spacecraft.

There is no obvious unique time interval to expect between intermittent signals or to suggest their duration, but the planetary day is one time interval that would affect transmission and observing operations from the surface of rotating planets and might affect the interval between signals. In a general case, transmissions from a single site are possible only during the part of the day when targets are above the local horizon, causing a periodic daily absence of signal for some observers regardless of any other intended repetition rate. Similarly, observations of some targets are only possible for the portion of the day when the targets are above the horizon. In the special case of a fixed high-gain antenna system sweeping across the sky with the rotation of a planet, the signal would appear to be periodic with the source planet's day. Such fixed antenna systems might be necessary in signaling programs extending over extremely long periods of time.

The distribution of planetary days in our solar system provides a guide to days elsewhere. Two-thirds of the nine traditional planets have days in the range of approximately 10-25 hours (three-quarters with Pluto excluded), and the median day is 23.9 hours (21.1 hours with Pluto excluded). The two planets in or near the current solar habitable zone, Earth and Mars, have 24 and 25 hour days respectively, although the terrestrial day was 18 hours one billion years ago and will grow longer in the future. The range of 10-25 hours seems plausible as an estimate for the 'day' of many planets in other planetary systems.

A planetary day time scale might be useful in searching for interstellar signals, because planetary rotation would have physical effects on both transmissions and searches operating on the surface of planets. Observations over a planetary day would off course cover many possible shorter repetition rates; observations extending over approximately 25 hours would include signal repetition rates up to the 66th percentile of days in our solar system. That is a much longer observing time than is typical in SETI, but techniques such as radio imaging can be used to observe many stars in a wide field simultaneously. Observations over less than 10 hours would not cover even the shortest planetary days in our solar system.

Most targeted searches have assumed continuous signals, observing targets very briefly, in part due to the historical constraint of limited observing time. But, physical and economic constraints might make some or many signals intermittent, such as planetary rotation shadowing targets and limited resources resulting in low-duty-cycle isotropic or targeted transmissions. It is possible that there are no interstellar signals that are always present, so that searches will need to observe for extended periods of time (or large numbers of targets) in order to find and confirm interstellar signals if any exist.

## Acknowledgments

RG thanks James Benford, Steven Lord, Nick Oberg, and an anonymous referee for comments on the manuscript. The SAS Institute supported this work in part with a software license grant for the Statistical Analysis System.